\documentclass[aps,prl,twocolumn,superscriptaddress,floatfix]{revtex4-1}

\usepackage{graphicx}
\usepackage{amsmath,amssymb}
\usepackage{hyperref}
\usepackage[utf8]{inputenc}
\usepackage{lipsum}
\usepackage[normalem]{ulem}

\usepackage{graphicx}
\usepackage{dcolumn}
\usepackage{bm}
\usepackage{hyperref}
\usepackage[utf8]{inputenc}
\usepackage[english]{babel}
\usepackage{units}
\usepackage{mathtools}
\hypersetup{colorlinks=true,citecolor={blue},linkcolor={blue},urlcolor={blue}}
\usepackage{float}
\usepackage{amsmath}
\usepackage{empheq}
\usepackage{mathrsfs}
\usepackage{amssymb}
\usepackage{soul}
\usepackage{textcomp}
\usepackage{placeins}
\usepackage{xcolor}
\usepackage[T1]{fontenc}
\usepackage{tipa}
\setstcolor{red}

\begin{document}

\title{Coherence Revivals and Lifetime Extension of Polariton Condensates by Mirror-Mediated Self-Feedback}

\author{I. Smirnov}
\affiliation{Skolkovo Institute of Science and Technology, Territory of innovation center “Skolkovo”, Bolshoy Boulevard 30, bld. 1, 121205 Moscow, Russia}

\author{S.~Alyatkin}
\affiliation{Skolkovo Institute of Science and Technology, Territory of innovation center “Skolkovo”, Bolshoy Boulevard 30, bld. 1, 121205 Moscow, Russia}

\author{P.~G.~Lagoudakis}
\affiliation{Skolkovo Institute of Science and Technology, Territory of innovation center “Skolkovo”, Bolshoy Boulevard 30, bld. 1, 121205 Moscow, Russia}

\date{\today}

\begin{abstract}
Temporal coherence of driven-dissipative condensates is limited by phase noise. We show that mirror-mediated time-delayed self-feedback enables control of coherence in a trapped exciton-polariton condensate. Reinjecting a small fraction of the emitted light with a tunable delay reveals two regimes set by the ratio of delay time to intrinsic coherence time. Long delays result in pronounced coherence revivals at integer multiples of the feedback delay, while short delays suppress phase diffusion and nearly double the coherence time. A minimal stochastic delayed model reproduces both regimes and supports an interpretation in terms of phase stabilization and delay-induced spectral filtering.

\end{abstract}

\maketitle
Microcavity polaritons represent a hybrid matter–wave system that under the optical drive can form macroscopically coherent states called polariton condensates~\cite{Kasprzak2006,Deng2010}. Their driven-dissipative nature, combined with intrinsic nonlinearity, provides direct optical access to excitation, manipulation and reading out of their states~\cite{Carusotto2013}.
Moreover, the macroscopic coherence of polariton condensates naturally emerges in the presence of gain, loss, interactions, and noise. The nonequilibrium character of the polariton system makes temporal coherence both a fundamental observable and a key parameter for analogue simulation and photonic computing platforms~\cite{Amo2016,Boulier2020,Kavokin2022}. So far, the coherent properties and synchronization of polariton condensates have been investigated in their clusters~\cite{Tosi2012,Ohadi2016,Cookson2021} and lattices of almost arbitrary geometries~\cite{Kim2011_square,Masumoto2012_kagome,Berloff2017,Whittaker_Lieb,Alyatkin_APL2024,Alyatkin2025_SciAd}. Considerable progress has been made in extending temporal coherence in confined condensates~\cite{Klaas2018,Orfanakis2021,Fabricante24}, engineering spatial coherence~\cite{Topfer2021}, as well as control of the node-to-node coupling strength~\cite{Ohadi2017}, its directionality~\cite{Wang2022_PRB} and phase~\cite{Ohadi2016,Alyatkin_PRL}. At the same time, finite propagation and feedback times make delayed interactions an intrinsic ingredient of polariton systems~\cite{Topfer2020}, connecting them to the broader physics of time-delay dynamics and non-Markovian networks~\cite{Richard2003,Lin2020}.

Despite recent advances in fundamental studies and technologies, developing simple and tunable strategies for controlling the coherence of polariton condensates in the presence of strong nonlinearities and dissipation~\cite{Byrnes2014,Comaron2025} represents a central challenge for real-world applications. An alternative route to coherence control in open oscillators is time-delayed self-feedback, in which a fraction of the emitted field is reinjected after a controlled delay. In polariton condensates, mirror-mediated optical feedback has recently been shown to enhance emission and reduce the threshold through delayed seeding in pulsed excitation~\cite{Mirek2025}, and to generate programmable long-range coupling and phase locking between spatially separated condensates~\cite{Liang2025}. 

\begin{figure}[t]
  \centering
  \includegraphics[width=\columnwidth]{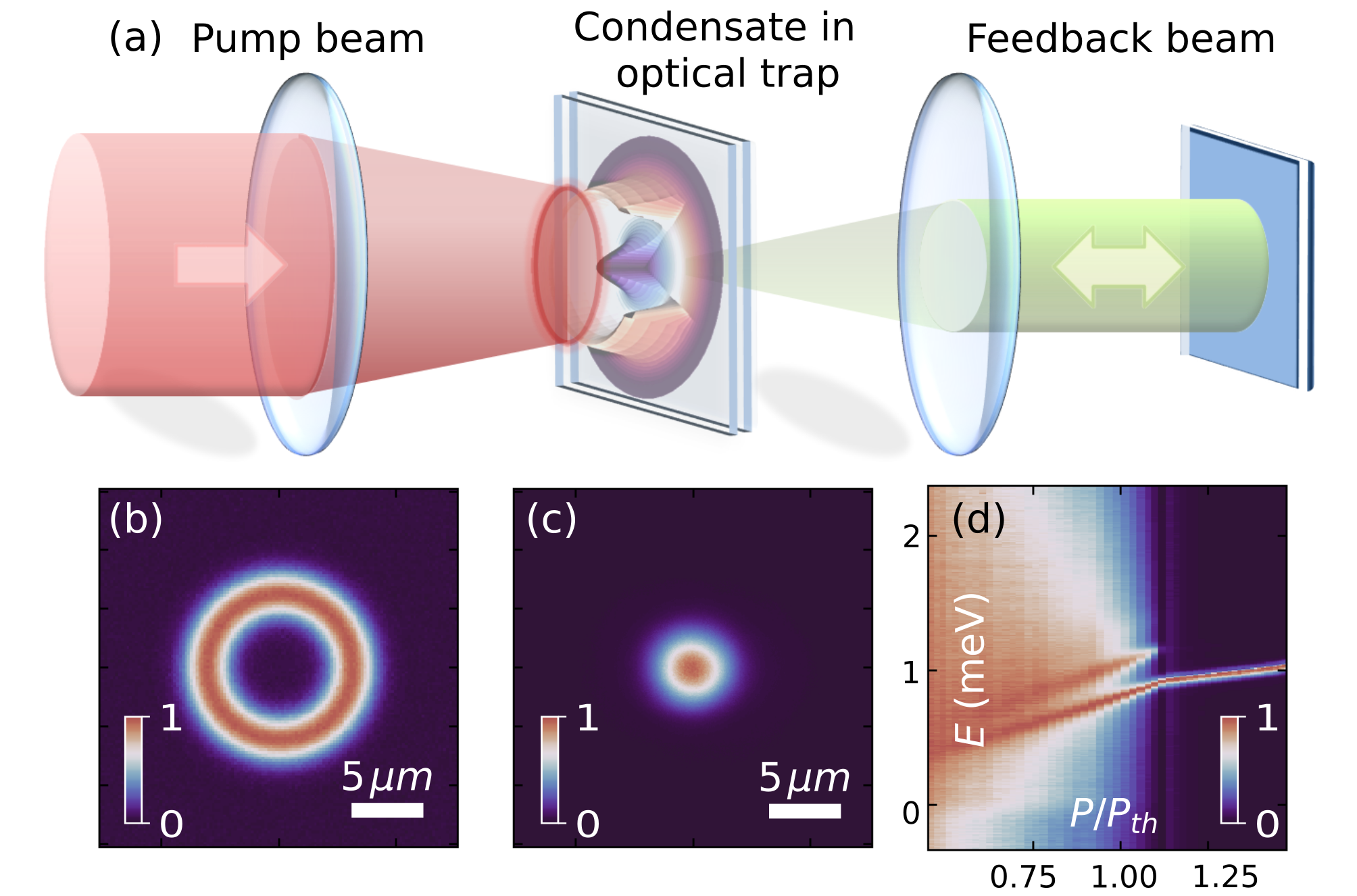}
  \caption{(a) Schematic of the experiment: a trapped polariton condensate is coupled to its time-delayed replica by retroreflection from an external mirror placed at distance $L$, corresponding to a round-trip delay $\tau_F=2L/c$. (b) Real-space intensity profile of the annular nonresonant pump. (c) Time-averaged real-space photoluminescence (PL) of the trapped condensate at $P=1.2P_{\mathrm{th}}$. (d) Normalized PL spectrum versus pump power, showing the formation of a localized single-mode condensate above threshold. Energies are referenced to the bottom of the lower polariton branch.}
  \label{fig1}
\end{figure}

In this Letter, we demonstrate that time-delayed self-feedback provides a simple approach for node-level coherence control in a single continuously-driven trapped polariton condensate. Treating an optically confined condensate as a minimal platform for delayed back-action [Fig.~\ref{fig1}(a)], we independently vary the feedback time-delay $\tau_F$ and reinjection feedback $\delta$, and measure the first-order coherence function $g^{(1)}(\tau)$. We find that the ratio $\tau_F/\tau_c$, with $\tau_c$ the intrinsic coherence time, selects between two distinct regimes. For long delays ($\tau_F \gtrsim \tau_c$) the condensate exhibits pronounced coherence revivals at integer multiples of the delay, while for short delays ($\tau_F < \tau_c$) feedback suppresses phase diffusion and substantially extends the coherence time without resolved beatings. A simple stochastic delayed model captures both regimes, explaining them through phase stabilization and delay-induced spectral selection. Our results identify self-feedback as a practical tool for non-Markovian control of macroscopic coherence in polariton condensates.

The experiments are performed on a strain-compensated planar GaAs-based semiconductor microcavity held at $T \approx 4$ K~\cite{Cilibrizzi2014}. We nonresonantly excite the sample with a continuous-wave laser at 796 nm and an exciton-photon detuning of $-4.3$ meV, see Supplemental Material, Section S1 (SM~S1) for further information on the microcavity structure. The pump is shaped into an annular profile to create a $10~\mu\mathrm{m}$-diameter optical trap [Fig.~\ref{fig1}(b)] for a ground-state localized condensate [Fig.~\ref{fig1}(c)]. Above threshold pump $P_{th}$, a single-mode polariton condensate is emerging, as shown in the time-integrated $P/P_{th}$ spectral dependence of Fig.~\ref{fig1}(d). The trapped condensate has a reduced overlap with the incoherent exciton reservoir, thereby mitigating reservoir-induced dephasing~\cite{askit_arxiv}. We implement mirror-mediated self-feedback by retroreflecting a controlled fraction of the condensate emission back onto the microcavity [Fig.~\ref{fig1}(a)], so that the condensate field $\psi(t)$ is coupled to its delayed replica $\psi(t-\tau_F)$, with round-trip delay $\tau_F$. Temporal coherence is measured with a Michelson interferometer through the decay time $\tau$ dependent fringe visibility, from which we extract the first-order coherence function $g^{(1)}(\tau) = \frac{\langle \psi^\ast(t)\psi(t+\tau)\rangle}{\langle|\psi(t)|^2\rangle}$, see SM S2 for experimental methods.

\begin{figure}[t]
  \centering
  \includegraphics[width=\columnwidth]{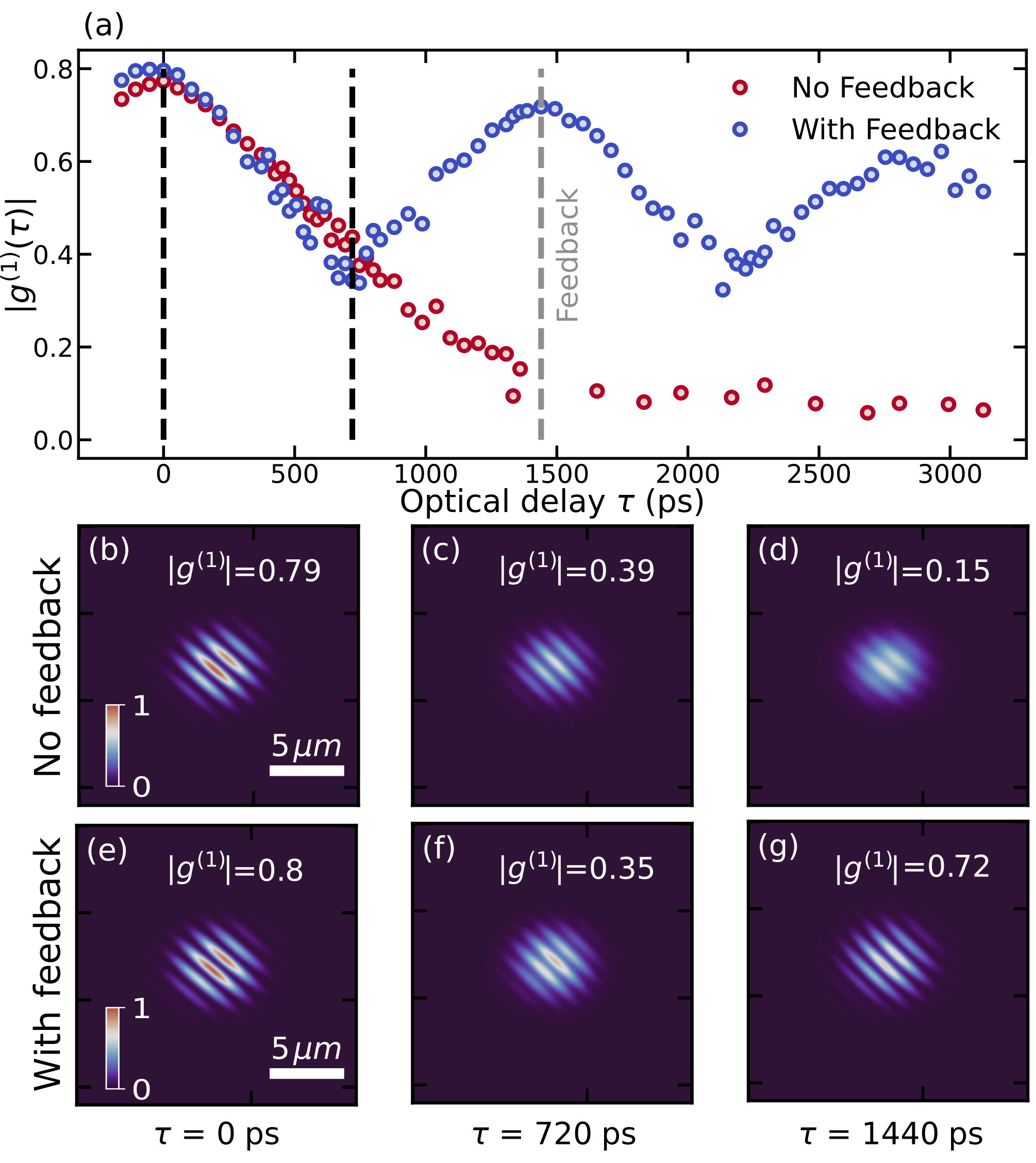}
 \caption{Long-delay feedback regime, $\tau_F \gtrsim \tau_c$. (a) Measured $|g^{(1)}(\tau)|$ for the trapped polariton condensate at $P \approx 1.2P_{\mathrm{th}}$ without feedback (red circles) and with delayed self-feedback (blue circles) for $\tau_F=1.44$ ns and estimated feedback rate $\delta\approx 2.8\%$. Dashed vertical lines mark $\tau=0$, $\tau_F/2$, and $\tau_F$. (b)-(d) Measured single-shot interferograms without feedback at the indicated delays, showing intrinsic coherence decay. (e)-(g) Corresponding interferograms with feedback, showing coherence revival when the delayed field returns to the condensate.}
  \label{fig2}
\end{figure}

\begin{figure}[t]
  \centering
  \includegraphics[width=\columnwidth]{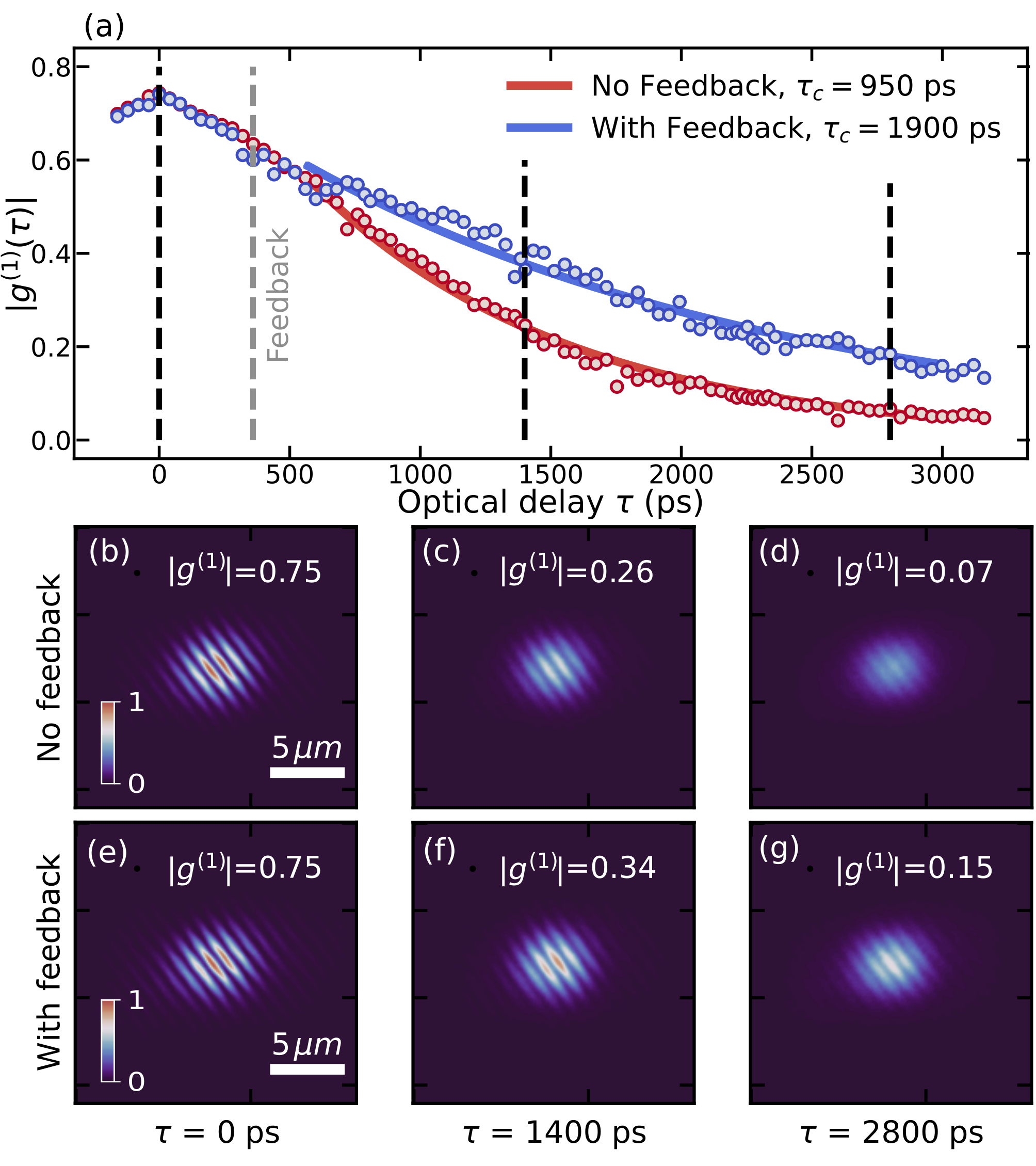}
\caption{Short-delay feedback regime, $\tau_F < \tau_c$. (a) Measured $|g^{(1)}(\tau)|$ for the trapped polariton condensate at $P \approx 1.4P_{\mathrm{th}}$ without feedback (red circles) and with delayed self-feedback (blue circles) for $\tau_F=350$ ps and estimated feedback rate $\delta\approx 2.2\%$. (b)-(d) Single-shot interferograms without feedback at the indicated delays, showing intrinsic coherence decay. (e)-(g) Corresponding interference patterns with feedback, showing an extended coherence time without resolved revivals.}
  \label{fig3}
\end{figure}

We first consider the long-delay regime, $\tau_F \gtrsim \tau_c$. Figure~\ref{fig2} compares $|g^{(1)}(\tau)|$ without feedback and with self-feedback at $P = 1.2P_{\mathrm{th}}$ and $\tau_F = 1.44$ ns, with an estimated $\delta\approx 2.8\%$, see SM~S3. Without feedback, $|g^{(1)}(\tau)|$ decays approximately exponentially with $\tau_c \approx 1$ ns, as shown with red circles in Fig.~\ref{fig2}(a), consistent with earlier measurements of trapped condensates~\cite{askit_arxiv,Orfanakis2021,Sigurdsson2022_PRL}. However, with feedback applied, the behaviour changes qualitatively: $|g^{(1)}(\tau)|$ develops pronounced revivals (see blue circles in Fig.~\ref{fig2}(a)) at integer multiples of $\tau_F$, and at $\tau = \tau_F$ reaches $\sim 0.8\,|g^{(1)}(0)|$, far exceeding the intrinsic decay. The measured  single-shot interference patterns in Figs.~\ref{fig2}(b)--~\ref{fig2}(g) show the same physics directly in real space: coherence is lost over the intrinsic decay time in the absence of feedback, but is restored when the delayed field returns to the condensate. These measurements establish delayed phase-relocking as the characteristic signature of the long-delay regime.

We next turn to the short-delay regime, $\tau_F < \tau_c$. Figure~\ref{fig3} compares $|g^{(1)}(\tau)|$ without feedback and with self-feedback at $P = 1.4P_{\mathrm{th}}$ and $\tau_F = 350$ ps, with an estimated $\delta\approx 2.2\%$, see SM~S3. In marked contrast to Fig.~\ref{fig2}, the delayed feedback does not generate resolvable revivals in $|g^{(1)}(\tau)|$. Instead, it slows the coherence decay, increasing the coherence time by nearly a factor of two. The interference patterns in Figs.~\ref{fig3}(b)--~\ref{fig3}(g) confirm that the major relevant effect in this regime is not delayed recovery after coherence is lost, but extended phase coherence. 

\begin{figure}[t]
  \centering
  \includegraphics[width=\columnwidth]{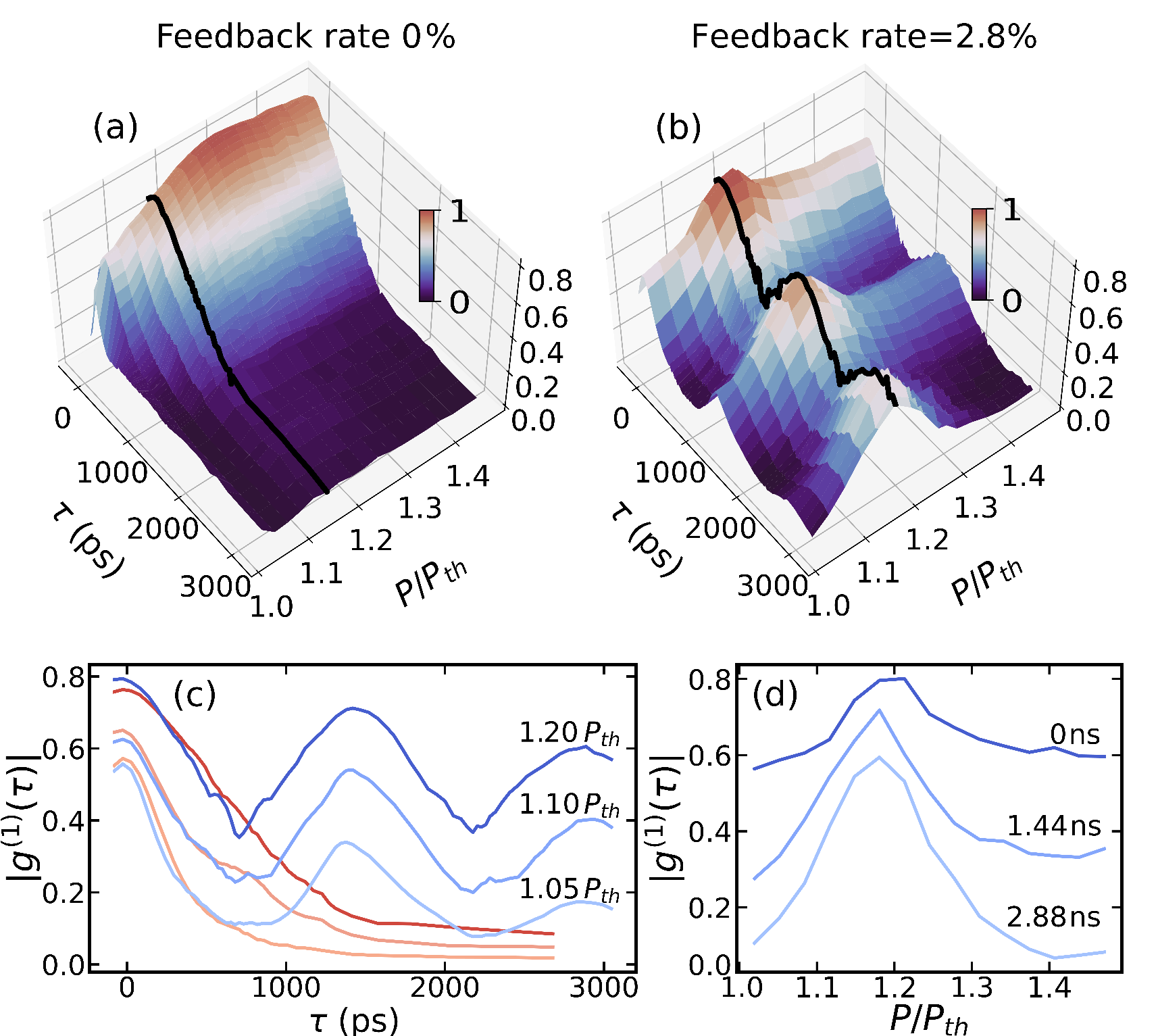}
\caption{Pump-power dependence of temporal coherence in the long-delay regime. (a),(b) Measured $|g^{(1)}(\tau)|$ as a function of pump power and optical delay without feedback and with delayed self-feedback, respectively, for $\tau_F=1.44$ ns and $\delta\approx 2.8\%$. (c) Cross-sections of the maps in (a) and (b) at fixed pump power, showing the evolution of $|g^{(1)}(\tau)|$ with $P$. (d) Cross-sections at fixed delay, $\tau=0$, $\tau_F$, and $2\tau_F$, highlighting the pump dependence of the revival amplitude. Black curves in (a) and (b) reproduce the data shown in Fig.~\ref{fig2}(a).}
  \label{fig4}
\end{figure}

Further on we investigate the broader experimental map of $|g^{(1)}(\tau)|$ versus pump $P/P_{th}$ and decay time $\tau$, that also justifies the choice of operating conditions under which the coherence revivals of Fig.~\ref{fig2} were presented. The upper colour-plots compare the system without and with feedback at $\tau_F = 1.44$~ns and $\delta\approx 2.8\%$. Figure~\ref{fig4}(a) demonstrates that in the absence of feedback, increasing the pump from $1.05 P_{\rm th}$ to $1.3 P_{\rm th}$ results in a growth of $|g^{(1)}(\tau)|$ across all $\tau$, with maximum value $|g^{(1)}(0)| \approx 0.85$ and coherence time of $\tau_c \approx 1$~ns, compared to $\tau_c \approx 200$~ps at $P_{th}$. At higher pump, up to $1.5 P_{\rm th}$, the coherence remains essentially unchanged~\cite{Sigurdsson2022_PRL}. 
In the presence of feedback, the pump dependence becomes non-trivial, as visible from Fig.~\ref{fig4}(b). The observed coherence revivals become especially pronounced at $\tau \approx \tau_F$ at $P\approx1.2 P_{\rm th}$, with extracted peak value almost approaching the magnitude $|g^{(1)}(0)|$, decreasing abruptly for both lower and higher $P/P_{th}$. Cross-sections at fixed $P$ in Fig.~\ref{fig4}(c) illustrate the temporal evolution of $|g^{(1)}(\tau)|$, while cross-sections at fixed $\tau$ in Fig.~\ref{fig4}(d) reveal the existence of well-defined pumping excitation conditions for the emergence of the coherence revivals. 

\begin{figure}[t]
  \centering
  \includegraphics[width=\columnwidth]{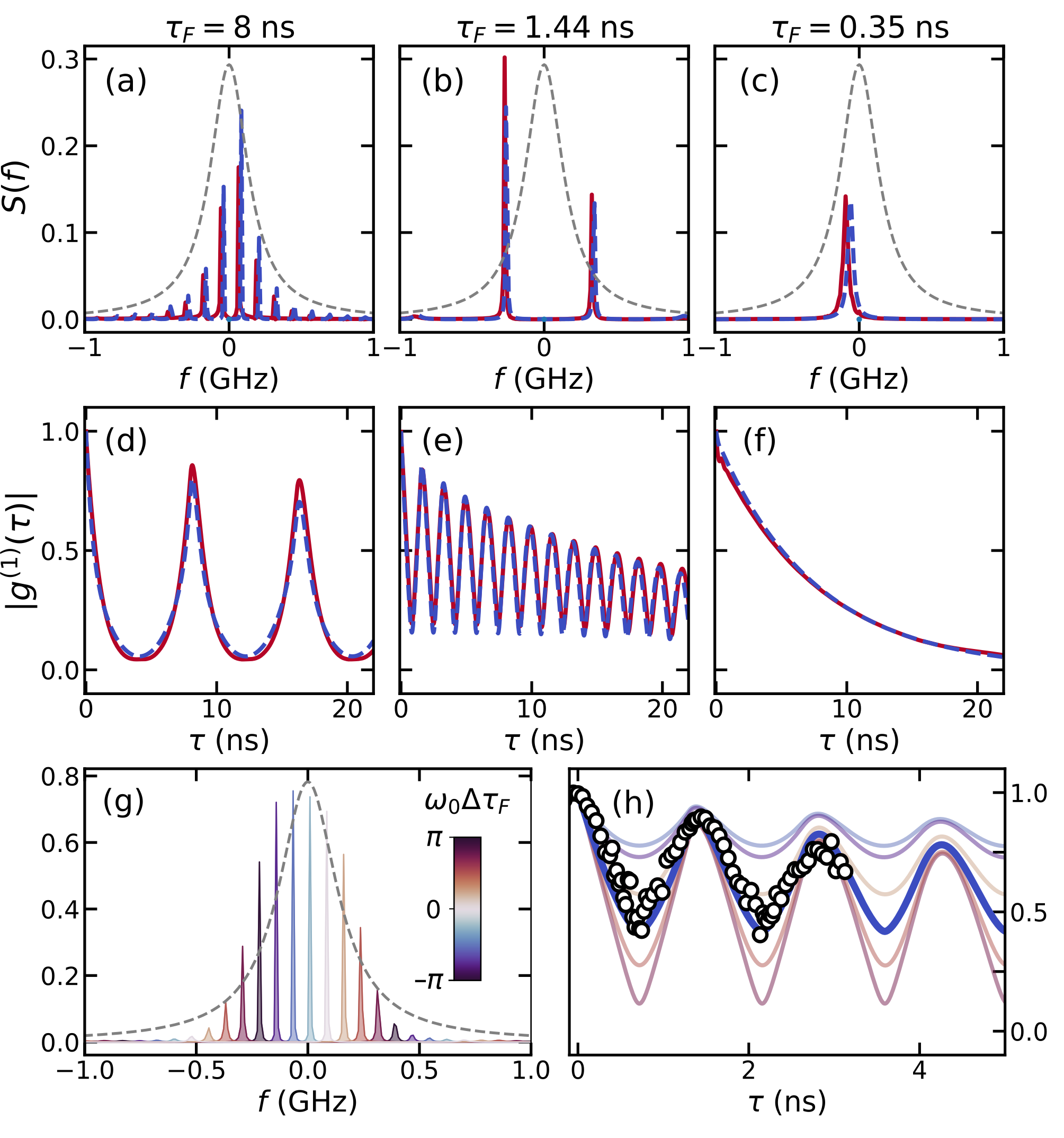}
\caption{Delayed-feedback model: numerical solution of Eq.~\eqref{eq:GPE_OD} (red) and analytical result from Eq.~\eqref{eq:spec_dens} (blue). (a)--(c) Simulated spectra for different delay times $\tau_F$. Gray dashed curves indicate the intrinsic condensate linewidth without feedback. (d)--(f) Corresponding $|g^{(1)}(\tau)|$. (g) Dependence of the spectrum on the feedback phase $\omega_0\tau_F$. (h) Individual theoretical realizations of $|g^{(1)}(\tau)|$ for different feedback phases (colored curves), their phase-averaged profile (blue), and the normalized experimental data from Fig.~\ref{fig2} (black circles).}
  \label{fig5}
\end{figure}

Figures~\ref{fig2}–\ref{fig4} show that the same delayed self-feedback can either restore phase correlations at discrete delays or extend coherence monotonically, depending on the ratio $\tau_F / \tau_c$. To capture both regimes within a single framework, we model the trapped condensate as a zero-dimensional order parameter $\psi(t)$. This reduction is justified by the annular trap, which supports a localized single-mode condensate with weak reservoir overlap. The dynamics is described by the driven-dissipative Gross-Pitaevskii equation with delayed reinjection:
\begin{equation*}
i\dot{\psi}(t)=
\left[i(P_{\mathrm{eff}}-\gamma_c)+\omega_0+\alpha|\psi(t)|^2\right]\psi(t)
\end{equation*}
\begin{equation}
+i\delta\gamma_c\psi(t-\tau_F)+\xi(t)
\label{eq:GPE_OD}
\end{equation}
where $P_{\mathrm{eff}}=P/(1+\zeta|\psi|^2)$ is the saturable pump, $\gamma_c$ is the cavity loss rate, $\alpha$ is the interaction constant, $\delta \in [0,1]$ corresponds to a fraction of the reinjected polariton field, and $\xi(t)$ is complex Gaussian noise such that $\langle\xi(t)\rangle = 0$ and $\langle\xi(t)\xi^*(t')\rangle = D_0\delta(t-t')$. Equation~\eqref{eq:GPE_OD} is used to numerically simulate the condensate dynamics (see Fig.~\ref{fig5} and SM S4). We now assume that coherence is dominated by phase rather than intensity fluctuations and derive an analytical description (see SM S5). We neglect intensity fluctuations and retain only fluctuations of the complex field around the stationary state. In this approximation, $|\psi(t)|^2\simeq I_s$ is fixed, so that the nonlinear frequency shift is absorbed into the blue-shifted frequency $\tilde{\omega}_0=\omega_0+\alpha I_s$. The dynamics then reduce to the linear delayed Langevin equation:
\begin{equation}
i\dot{\psi}(t)=\bigl[\tilde{\omega}_0-i\gamma_{\mathrm{eff}}\bigr]\psi(t)
+i\delta\gamma_c\,\psi(t-\tau_F)+\xi(t),
\label{eq:linear_delay}
\end{equation}
where $\gamma_{\mathrm{eff}}$ is an effective damping parameter accounting for the residual gain-loss balance in the reduced description, whose value in the absence of feedback can be estimated by the measured $\tau_c$. In the presence of time-delayed feedback we find that $\tau_c$ scales linearly with $\delta$, which  consequently affects $\gamma_{\mathrm{eff}}$, see SM S4.

Equation~\eqref{eq:linear_delay} admits an exact solution, see SM S5, from which the spectral density follows as
\begin{equation}
S(\omega)=
\left|
\frac{1}{i(\tilde{\omega}_0-\omega)+\gamma_{\mathrm{eff}}-\gamma_c\delta e^{i\omega\tau_F}}
\right|^2
\label{eq:spec_dens}
\end{equation}
In the absence of feedback, Eq.\eqref{eq:spec_dens} reduces to a single Lorentzian, corresponding to the exponential decay of $g^{(1)}(\tau)$ in agreement with the Wiener–Khinchin theorem. Finite feedback generates a ladder of delay-induced resonances. Crucially, only resonances that fall within the intrinsic condensate linewidth contribute significantly, so delayed self-feedback acts as a spectral-selection mechanism controlled by $\tau_F$ and $\delta$.

When $\tau_F \gg \tau_c$, Eq.~\eqref{eq:spec_dens} yields a comb of narrow resonances superimposed on the intrinsic condensate linewidth [gray curve in Fig.~\ref{fig5}(a)]. In the time domain this corresponds to revivals of $|g^{(1)}(\tau)|$ at integer multiples of $\tau_F$, separated by intervals of nearly complete coherence loss [Fig.~\ref{fig5}(d)]. As $\tau_F$ is reduced toward the intrinsic coherence time, the spacing between resonances increases and only a few modes remain within the intrinsic linewidth. For the experimentally relevant case $\tau_F=1.44$ ns, this leaves only two dominant spectral peaks [Fig.~\ref{fig5}(b)], which account for both the enhanced coherence and the oscillatory structure of $|g^{(1)}(\tau)|$ in Fig.~\ref{fig5}(e), in agreement with Fig.~\ref{fig2}(a). In the short-delay regime, $\tau_F<\tau_c$, the resonances are pushed outside the intrinsic linewidth and the spectrum is effectively single-mode [Fig.~\ref{fig5}(c)]. Accordingly, $|g^{(1)}(\tau)|$ recovers a monotonic, approximately exponential decay [Fig.~\ref{fig5}(f)], but with a substantially increased coherence time, consistent with Fig.~\ref{fig3}(a).

Fluctuations of the feedback phase provide an additional source of broadening in the experiment. Since each realization corresponds to a different effective phase $\omega_0\tau_F$, the measured coherence should be compared with the phase-averaged prediction of Eq.~\eqref{eq:spec_dens}, obtained by averaging over $\delta\rightarrow \delta e^{i\phi}$ with $\phi\in[0,2\pi]$. Figure~\ref{fig5}(g) shows the corresponding modification of the spectrum, while Fig.~\ref{fig5}(h) compares individual realizations, their average, and the normalized experimental data from Fig.~\ref{fig2}. The agreement captures not only the mean revival pattern but also the enhanced spread at the minima of $|g^{(1)}(\tau)|$ and the reduced spread at the maxima.

These results show that delayed self-feedback acts as a spectral-selection mechanism: only resonances supported by the delayed cavity and lying within the intrinsic condensate linewidth contribute appreciably to the long-time coherence. In this sense, Eq.~\eqref{eq:spec_dens} provides a simple physical picture for both observed regimes. For long delays, multiple resonances participate and generate revivals; for short delays, only a single resonance remains relevant, yielding linewidth narrowing without resolved beatings. The analogy to a Fabry-P\'erot filter is useful at the spectral level, although the condensate dynamics remain intrinsically nonlinear and noisy.

In conclusion, we have shown that mirror-mediated time-delayed self-feedback provides a simple and tunable route to coherence control in a trapped exciton-polariton condensate. By varying the ratio $\tau_F/\tau_c$, we identify two qualitatively distinct regimes: delayed phase relocking and coherence revivals for $\tau_F\gtrsim \tau_c$, and linewidth narrowing with nearly twofold coherence-time enhancement for $\tau_F<\tau_c$. The developed stochastic delayed model reproduces experimental observations and attributes them to delay-induced spectral selection together with phase stabilization. More broadly, our results establish self-feedback as a practical mechanism for controlling non-Markovian coherence in polariton platforms and for engineering programmable out-of-plane couplings in delay-coupled condensate networks.

\begin{acknowledgments}
This work was supported by the Russian Science Foundation (RSF), grant No.~24-72-10118, \url{https://rscf.ru/en/project/24-72-10118/}.
\end{acknowledgments}

\bibliography{ref.bib}

\setcounter{equation}{0}

\setcounter{figure}{0}

\setcounter{section}{0}

\setcounter{subsection}{0}

\newcolumntype{P}[1]{>{\centering\arraybackslash}p{#1}}

\newcolumntype{M}[1]{>{\centering\arraybackslash}m{#1}}

\renewcommand{\theequation}{S\arabic{equation}}

\renewcommand{\thefigure}{S\arabic{figure}}

\renewcommand{\baselinestretch}{1}

\onecolumngrid

\newpage

\vspace{1cm}

\begin{center}

\Large \textbf{Supplementary Information}

\end{center}


\section{MICROCAVITY STRUCTURE}

The experiments were performed on a strain-compensated planar GaAs-based semiconductor microcavity described in Ref.~\cite{Cilibrizzi2014}. The structure consists of a $2\lambda$ GaAs cavity bounded by distributed Bragg reflectors formed by alternating GaAs and AlAs$_{0.98}$P$_{0.02}$ layers, with 26 pairs in the bottom mirror and 23 pairs in the top mirror. Three pairs of $6~\mathrm{nm}$ In$_{0.08}$Ga$_{0.92}$As quantum wells are embedded at the antinodes of the cavity field, together with two additional quantum wells placed at the outer nodes to improve carrier collection. The strain-compensated design reduces structural disorder, while the cavity wedge enables tuning of the exciton--photon detuning. In the present work, the sample was held at $T \approx 4~\mathrm{K}$ and operated at exciton--photon detuning $\Delta=-4.3~\mathrm{meV}$.

\section{EXPERIMENTAL METHODS}

The sample was nonresonantly excited with a laser at $\lambda=796~\mathrm{nm}$. The pump was shaped into an annular profile of diameter $\sim 10~\mu\mathrm{m}$, forming an optical trap that supports a localized single-mode condensate with reduced overlap with the incoherent exciton reservoir.The condensate emission was analyzed in real space and in the energy domain. Interferometric measurements were performed using Michelson interferometer. The first-order coherence function $g^{(1)}(\tau)$ was extracted from the delay-dependent fringe visibility.

For time-averaged measurements, the excitation was intensity-modulated at $5~\mathrm{kHz}$ with a duty cycle of $1\%$. In this regime, we acquired the time-averaged real-space photoluminescence and pump-dependent spectra used to characterize condensate formation [Figs.~1(c) and 1(d)].

For single-shot measurements, the excitation pulse duration was set to $35~\mu\mathrm{s}$. The exposure time was $35~\mathrm{ms}$ and acquisition was synchronized with the pump signal to ensure that each frame captured a single condensate realization. The short excitation duration further reduces the influence of mechanical vibrations in the optical setup, while remaining sufficient to acquire statistics of the condensate temporal dynamics. In this regime, we performed delay-dependent interferometric measurements to extract $|g^{(1)}(\tau)|$. The corresponding data are presented in Figs.~2 and 3. The same acquisition mode was employed to obtain the two-dimensional maps of $|g^{(1)}(\tau)|$ as a function of pump power and optical delay, shown in Fig.~4.

\section{FEEDBACK RATE ESTIMATION}

In this section, we estimate the effective feedback rate $\delta$ by accounting for the dominant optical losses in the feedback path. The total reduction of the reinjected field has two main contributions: losses in the external optical elements and the transmission/reflection properties of the microcavity itself.

We first consider the long-delay configuration. The transmission of the collecting microscope objective is approximately $t_{\mathrm{obj}}=0.7$, the transmission of the cryostat windows is $t_{\mathrm{window}}=0.87$, and the reflectivity of the external feedback mirror is $r_{\mathrm{FM}}=0.95$. The intensity returning to the sample is therefore reduced to
\begin{equation}
I_F = I_0\, t_{\mathrm{obj}}^2 t_{\mathrm{window}}^2 r_{\mathrm{FM}}
      \equiv t_{\mathrm{total}}^2 I_0
      \approx 0.35\, I_0,
\end{equation}
where $I_0$ is the detected condensate PL intensity and $t_{\mathrm{total}}^2 \approx 0.35$ is the total transmission of the external optical path.

This estimate can be checked directly in the experiment. By deliberately misaligning the feedback spot with respect to the condensate, the two spots become spatially separated in real space. In this configuration, the blue-shifted condensate PL is fully reflected by the DBR, allowing us to compare the two intensities directly. We measure a feedback-spot intensity of approximately $0.35\,I_0$, in good agreement with the estimate above.

A second reduction arises from the planar microcavity. Since the feedback field is reinjected from the substrate side, we model the cavity using the transfer-matrix method (TMM). This allows us to account for the asymmetry between the two DBRs and for the coupling efficiency of an incident resonant field into the cavity. Owing to the asymmetric cavity design, the emission rates toward the reflection side and transmission side are different. For the present structure, with 23 pairs in the top DBR and 26 pairs in the bottom DBR, we obtain
\begin{equation}
\frac{\gamma_{\mathrm{ref}}}{\gamma_{\mathrm{tr}}} \approx 1.5,
\end{equation}
which contributes a reduction factor
\begin{equation}
\frac{1}{1+\gamma_{\mathrm{ref}}/\gamma_{\mathrm{tr}}} \approx 0.4.
\end{equation}

The TMM analysis also yields the coupling efficiency of the incident field into the cavity. For an intracavity source producing an outgoing intensity $I_0$, an incident beam with the same intensity generates an intracavity field amplitude
\begin{equation}
|E_F| = \eta |E_0| = 0.44\, |E_0|,
\end{equation}
where the reduction is dominated by reflection at the air--substrate interface.

Combining these factors, the effective feedback rate in the long-delay configuration is
\begin{equation}
\delta
= \frac{I_{\mathrm{returned}}}{I_{\mathrm{emitted}}}
= \eta^2
  \frac{1}{1+\gamma_{\mathrm{ref}}/\gamma_{\mathrm{tr}}}
  t_{\mathrm{total}}^2
\approx 2.8\%.
\end{equation}

For the short-delay configuration, we keep the same cavity-dependent factors, namely $\eta$ and $1/(1+\gamma_{\mathrm{ref}}/\gamma_{\mathrm{tr}})$. The external optical transmission is determined experimentally in the same way as above and yields
\begin{equation}
I_F \approx 0.28\, I_0.
\end{equation}
We therefore take $t_{\mathrm{total}}^2 = 0.28$ for the short-delay configuration. The reduced value, compared with the long-delay geometry, is attributed to imperfect focusing in the short-delay setup. This gives
\begin{equation}
\delta
= \eta^2
  \frac{1}{1+\gamma_{\mathrm{ref}}/\gamma_{\mathrm{tr}}}
  t_{\mathrm{total}}^2
\approx 2.2\%.
\end{equation}

\section{FEEDBACK RATE DEPENDENCE}

In this section, we summarize the dependence of the coherence properties on the feedback rate $\delta$, as obtained from numerical simulations of the delayed condensate dynamics.

For the numerical analysis, we fixed the feedback rate and feedback delay time and solved Eq.\eqref{eq:S1} to obtain time traces of $\Psi(t)$. The first-order coherence function was then evaluated directly from these data according to its definition. Each simulation was performed over a total time of $10~\mu\mathrm{s}$ with a temporal resolution of $5~\mathrm{ps}$, and the resulting coherence functions were averaged over $400$ realizations. The simulation parameters are as follow: $\gamma_c = 1/5.5 ~\mathrm{ps}^{-1}, \hbar\alpha = 1~\mathrm{\mu eV} , P_{\mathrm{eff}}=0.24 ~\mathrm{ps}^{-1}$, which corresponds to $ \approx 1.3 P_{\mathrm{th}}$. Stochastic noise ${\xi}(\omega) \sim \alpha_{noise}\mathcal{N}(0,1)$, where $\alpha_{noise} = 0.05 ~\mathrm{ps}^{-1}$ and $\mathcal{N}(0,1)$ denotes a normally distributed random variable with zero mean and unit variance.

For the analytical approach, we employed the Wiener–Khinchin theorem. Specifically, we numerically evaluated and normalized the spectral density given by Eq.\eqref{eq:S13}, followed by a Fourier transform to obtain $g^{(1)}(\tau)$. The coherence functions obtained using both methods are presented in Fig.~5 of the main text and are used throughout this section.

In order to avoid ambiguities in the fitting procedure of the $g^{(1)}(\tau)$ curve, we quantify the coherence enhancement using the ratio
\begin{equation}
\frac{|g^{(1)}(\tau_F)|}{|g^{(1)}(0)|},
\end{equation}
which directly probes the degree to which coherence is preserved over one feedback round trip.

\begin{figure}[t]
  \centering
  \includegraphics[width=0.4\columnwidth]{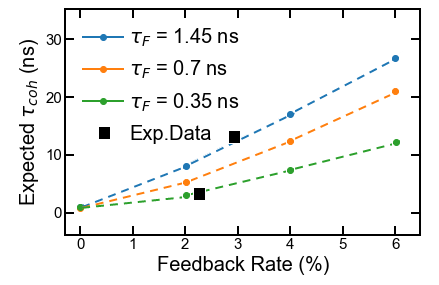}
 \caption{Dependence of the coherence indicator $|g^{(1)}(\tau_F)|/|g^{(1)}(0)|$ on the feedback rate $\delta$ for several delay times $\tau_F$, extracted from numerical simulations.}
  \label{fig1_SM}
\end{figure}

Figure~\ref{fig1_SM} shows the corresponding dependence for several delay times $\tau_F$. In all cases, the coherence increases approximately linearly with $\delta$ over the range considered here. The enhancement is strongest for longer delay times, for which the delayed field produces more pronounced coherence recovery. Within the reduced model, the coherence time can therefore be increased substantially by increasing the feedback rate.

At the same time, increasing $\delta$ also raises the condensate intensity. Experimentally, one may therefore expect that beyond some threshold the system will no longer exhibit a simple monotonic increase of coherence, but instead enter an unstable or chaotic regime. Such behavior is not captured by the present zero-dimensional model, which is intended only to describe the experimentally relevant weak-feedback regime.

\begin{figure}[t!]
  \centering
  \includegraphics[width=0.4\columnwidth]{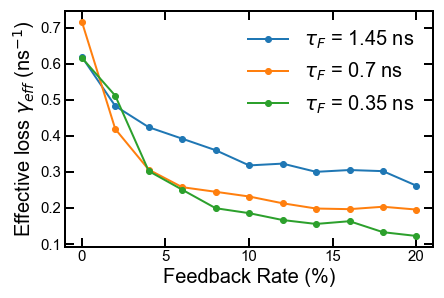}
 \caption{Dependence of the effective damping parameter $\gamma_{\mathrm{eff}}$ on the feedback rate $\delta$ for several delay times $\tau_F$, extracted from numerical simulations. The dominant linear contribution $\delta\gamma_c$, which reflects compensation of the feedback-induced gain in the resonant mode, is subtracted to present remaining nontrivial dependence.}
  \label{fig2_SM}
\end{figure}

Figure~\ref{fig2_SM} shows the dependence of the effective damping parameter $\gamma_{\mathrm{eff}}$ on the feedback rate. In the reduced linear description introduced in Sec.~S5, $\gamma_{\mathrm{eff}}$ is the only free parameter required to reproduce the numerically evaluated coherence dynamics. To extract this dependence, we solved the delayed condensate equation for a range of feedback rates $\delta$ and delay times $\tau_F$, computed the corresponding first-order coherence functions $g^{(1)}(\tau)$, and fitted them with the analytical expression derived in Sec.~S5. 

The dependence $\gamma_{\mathrm{eff}}(\delta)$ contains a dominant contribution proportional to $\delta \gamma_c$, which reflects compensation of the feedback-induced gain in the resonant mode. The remaining dependence $\hat\gamma_{\mathrm{eff}}=\gamma_{\mathrm{eff}}-\delta\gamma_c$ is plotted in Fig.\ref{fig2_SM}. It is nontrivial and encodes the residual effect of delayed reinjection on the effective linewidth within the reduced description.

\section{ANALYTICAL MODEL DERIVATION}

In this section, we derive the analytical expression for the spectral density used in the main text. We start from the stochastic delayed driven--dissipative Gross--Pitaevskii equation
\begin{equation}
i\dot{\psi}(t)=\left[i\left(P_{\mathrm{eff}}-\gamma_c\right)+\omega_0+\alpha |\psi(t)|^2\right]\psi(t)
+i\delta\gamma_c \psi(t-\tau_F)+\xi(t),
\label{eq:S1}
\end{equation}
with
\begin{equation}
P_{\mathrm{eff}}=\frac{P}{1+\zeta |\psi|^2}.
\label{eq:S2}
\end{equation}

To determine the stationary state, we first neglect noise and seek a monochromatic solution of constant intensity,
\begin{equation}
\psi_s(t)=\sqrt{I_s}\,e^{-i\omega_s t}.
\label{eq:S3}
\end{equation}
Using $\dot{\psi}_s=-i\omega_s\psi_s$ and $\psi_s(t-\tau_F)=\psi_s(t)e^{i\omega_s\tau_F}$, substitution of Eq.~(\ref{eq:S3}) into Eq.~(\ref{eq:S1}) gives
\begin{equation}
\omega_s=\omega_0+\alpha I_s+i\left[P_{\mathrm{eff}}(I_s)-\gamma_c\right]
+i\delta\gamma_c e^{i\omega_s\tau_F}.
\label{eq:S4}
\end{equation}
Separating real and imaginary parts, we obtain
\begin{align}
&\omega_s=\omega_0+\alpha I_s-\delta\gamma_c\sin(\omega_s\tau_F),
\label{eq:S5} \\
&P_{\mathrm{eff}}(I_s)-\gamma_c+\delta\gamma_c\cos(\omega_s\tau_F)=0.
\label{eq:S6}
\end{align}
Equation~(\ref{eq:S6}) determines the stationary intensity,
\begin{equation}
\frac{P}{1+\zeta I_s}
=\gamma_c\left[1-\delta\cos(\omega_s\tau_F)\right].
\label{eq:S7}
\end{equation}

We now neglect intensity fluctuations and retain only fluctuations of the complex field around the stationary state. In this approximation, $|\psi(t)|^2\simeq I_s$ is fixed, and the nonlinear blue shift is absorbed into
\begin{equation}
\tilde{\omega}_0=\omega_0+\alpha I_s .
\label{eq:S8}
\end{equation}
The dynamics then reduce to the linear delayed Langevin equation
\begin{equation}
i\dot{\psi}(t)=\left(\tilde{\omega}_0-i\gamma_{\mathrm{eff}}\right)\psi(t)
+i\delta\gamma_c\psi(t-\tau_F)+\xi(t),
\label{eq:S9}
\end{equation}
where we introduce an effective damping parameter $\gamma_{\mathrm{eff}}$ describing the residual gain--loss imbalance in the reduced model.

We use the Fourier-transform convention
\begin{equation}
\psi(t)=\int \frac{d\omega}{2\pi}\,\hat{\psi}(\omega)e^{-i\omega t}.
\label{eq:S10}
\end{equation}
Applying it to Eq.~(\ref{eq:S9}) yields
\begin{equation}
\hat{\psi}(\omega)=
\frac{\hat{\xi}(\omega)}
{i(\tilde{\omega}_0-\omega)+\gamma_{\mathrm{eff}}-\delta\gamma_c e^{i\omega\tau_F}}.
\label{eq:S11}
\end{equation}

Assuming white complex Gaussian noise,
\begin{equation}
\langle \hat{\xi}(\omega)\hat{\xi}^{*}(\omega')\rangle
=2\pi D_0\,\delta(\omega-\omega'),
\label{eq:S12}
\end{equation}
the spectral density of the condensate field is
\begin{equation}
S(\omega)\equiv \langle |\hat{\psi}(\omega)|^2\rangle
=
\frac{2\pi D_0}{
\left|i(\tilde{\omega}_0-\omega)+\gamma_{\mathrm{eff}}-\delta\gamma_c e^{i\omega\tau_F}\right|^2 }.
\label{eq:S13}
\end{equation}

Equation~(\ref{eq:S13}) is the spectral density used throughout the manuscript. Up to an overall normalization it reproduces Eq.~(3) of the main text.

\end{document}